\documentstyle[12pt]{article}
\medskip
\begin{document}
\pagestyle{empty}
\rightline{MRI-PHY/94/18}
          \begin{center}

     {\bf HAMILTONIAN PATH INTEGRAL QUANTIZATION IN ARBITRARY CO-ORDINATES
                              AND EXACT PATH INTEGRATION }
           \vspace{2cm}

                             ${\rm A.K. Kapoor}^ \dagger $

                               Mehta Research Institute,

                      10 Kasturba Gandhi Road (Old Kutcheri Road),

                                 Allahabad 211002, INDIA

           \vspace{2cm}
                                     Pankaj Sharan

                        Physics Department, Jamia Millia Islamia

                          Jamia Nagar, New Delhi 110025, INDIA

           \vspace{1.5cm}
                               \underline{ \bf ABSTRACT}
           \
           \end{center}
                We  briefly  review  a   hamiltonian   path  integral
formalism
                developed  earlier  by  one  of us. An important feature of
this
                formalism  is  that  the path integral quantization in
arbitrary
                co-ordinates   is   set   up   making   use  of  only
classical
                hamiltonian  without addition of  adhoc $\hbar^2$ terms. In
this
                paper  we use this hamiltonian formalism and show how exact
path
                integration may be done for several potentials.
                \vspace{1.0 cm}\\

                \hrule
                \noindent
$^\dagger $ Permanent Address : University of Hyderabad, Hyderabad 500134,
INDIA   \\
\ \  email : ashok@mri.ernet.in \\

           \pagebreak
\pagestyle{plain}

           \begin{center}
                           {\bf  1. Introduction }
           \end{center}

The Feynman path integral formalism has been in use for over forty years and
has  been  successfully  applied  to  a  variety of ${\rm problems}^1$.   An
important  development in the past fifteen years has been exact treatment of
large  number of potential problems within the path integral formalism. Duru
and  Kleinert   gave  an  exact path integral  treatment of  H-  atom  Green
function    using   Kustaanheimo-Stiefel   ${\rm  transformation}^3  $.  The
arguments  originally  used  by  Duru  and  Kleinert  were  formal  and  the
manipulations used lacked mathemaical justification. The H  atom problem was
given  a  correct  treatment  by  Ho  and  Inomata$^4$   soon after Duru and
Kleinerts  work.  It  has  since  been  possible   to   give an  exact  path
integral solution  of   several problems in quantum mechanics using the  new
technique   scaling  of local time. Notable  among  these  are  Morse  ${\rm
 oscillator}^5$,  Rosen-Morse  $ {\rm oscillator}^6 $,  Poschl-Teller  ${\rm
 potential} ^7$, Hartmann ${\rm potential}^8$, Hulthen ${\rm potential}^9$, $
\delta  $  - function   ${\rm potential}^{10}$,  square ${\rm well}^{11}$,
and  many  other  potential  problems  ${\rm  problem}^{12,13}$. Besides the
scaling  of  local  time  in  combination  with  change  of  variables,  the
techniques of adding new  degrees of freedom has been  useful in most of the
above mentioned applications.

\

  The path integral  formalism can be derived from quantum mechanics  and this
 aspect has  been a subject of extensive study in  the  literature.  One can
arrive at   different types  of  path  integral  representations  of  the
 propagator by working with different bases.  Thus,  for  example,  we
have  configuration space path integrals, phase space path integral  or
coherent state representation of path integrals. In all  these  derivations
quantum mechanics is assumed.

          \

One of the most important features of the path integral  formalism  is that it
 gives an alternative route to quantization.  In  use  of  the  path integrals
  as a scheme of quantization a prescription for  setting  up  path integral
  is assumed and quantum mechanics is  derived.  The  path  integral
  quantization  can  be  set  up  starting  from  classical   lagrangian   or
  hamiltonian  form of the classical action for the system of interest. In the
  hamiltonian formalism of classical  mechanics  one  can  write dynamical
  equations in any pair of conjugate variables related to  a  given set by a
  canonical transformation. However, it appears very difficult to implement
  {\it general} canonical transformations in  quantum  mechanics  even though
  there are a  large  number  of  investigations  on  this  $ {\rm
  subject^{14}} $. However, the point transformations do not present any
  difficulty in quantum mechanics. So it is of  interest  to  ask  if  path
  integral  quantization scheme, based on lagrangian or hamiltonian form  for
  the  action,  can  be formulated in arbitrary co-ordinates. The answer  has
  been  known  in  the lagrangian formulation known for a long time. The
  configuration space  path integral can  be  set  up  in  such  a  way  that
  it  works  in  arbitrary co-ordinates utilizing the configuration
  space form of classical ${\rm action}^{15}$.

                \

In a quantization scheme developed by one of ${\rm us} ^{16,17} $, quantization
is carried out in arbitrary co-ordinates using the hamiltonian  form  of  path
integrals. {\it A  unique feature of this  scheme  is  that  addition  of $
{O(\hbar^2 )} $   terms  is  not   necessary   in any   co-ordinate   system
and   only   the classical  hamiltonian  is used to set  up hamiltonian  path
integral quantization   in arbitrary   co-ordinates.} This formalism makes an
essential use of the  idea of local scaling of time to introduce a path
integral  representation  for quantum mechanical propagator.

                 \

In ref. 17 the basic property of the canonical  path  integrals,  that only
classical  hamiltonian  is  needed  for  quantization  in   arbitrary
co-ordinates, was established; discussion of  further properties and also of
applications was  not taken up. In ref.18 some important properties of  the
canonical path integral, such as scaling of local  time,  equivalence  with
lagrangian form of path integration have been investigated in detail. In
this paper possible applications  of  the  scheme  were  indicated  only
briefly. In the past fifteen years many exactly solvable quantum mechanical
problems have been treated by means of path  integration  by  relating  the
given problem to another problem for which exact path  integration  can  be
done
or is already known. In this paper we show  that  our  scheme   offers results
which can be used in a very simple manner to relate different  path integrals.
Although we, generally, follow the techniques already  available in the
literature to give exact treatment, the details  of  the  treatments are
different. There are some new points which require special and  careful
treatment.  For  example,  initial  condition  for   the propagator has to be
treated carefully.

                \

In the  next  section  we  summarize  all  the  necessary
results  on hamiltonian path integration from ref. 18.  Some  aspects  of
changes  of variables and the technique of adding new degrees of freedom are
discussed in Sec.3 and  Sec.4  respectively.  Application  of  the  hamiltonian
path integral method to  exact  solution  of  the  quantum  mechanical problems
is discussed in the Sec. 5 to 7.

\

\begin{center}
{\bf 2. Definition and properties of the hamiltonian path integrals}
\end{center}

In  this  section  we  briefly recall the definition of the hamiltonian path
${\rm  integral}^{16,17}$.  The  proofs   and  detailed  discussion  of  the
results can be found in our earlier ${\rm paper}^{18}$. We shall assume that
all   the hamiltonian like functions appearing in this paper are independent
of  time and are quadratic in momenta.

\

{\it 2.1 Definitions  }: Given a  phase  space  function $  H(q,p) $,  we
introduce a quantity $(qt \Vert q_0 t_0)$ which stands for the short time
approximation and  in terms of which a path integral for finite times is
constructed. We  assume $(qt \Vert q_0t_0)$ to be of the form

\begin{equation} (qt\Vert q_0 t_0) = \frac{1} {\sqrt{\rho (q)\rho (q_0)}} \int
d^n p_1 ~ (qt\vert p_1t_1) ~ (p_1 t_1\vert q_0 t_0)        \label{(2.1)}
\end{equation}

\noindent
The definition given below is such  that  (\ref{(2.1)})  satisfies  the  semi-
group property with respect to the measure $ \rho (q)d^n q $

\begin{equation} \int\rho(q_2) d^nq_2 (q_3 t_3 \Vert q_2 t_2) (q_2 t_2\Vert
q_1t_1)
 \approx  (q_3 t_3\Vert q_1t_1)         \label{(2.2)} \end{equation}

\noindent
when the integral is calculated in the stationary phase approximation. The
`mixed short time propagators $ (qt\vert p_1t_1)$ and $ (p_1t_1\vert q_0t_0)$
are defined below. Let $ \gamma_1 $ and $\gamma_2 $ be two classical
trajectories with boundary  conditions as indicated below.

\begin{equation} \gamma_1  : \tau   \rightarrow \bigl( \tilde {q}(\tau ),\tilde
{p}(\tau )\bigr)  ,\  \ {\rm for} \ \   t_0 \leq \tau  \leq t_1 \end{equation}

\begin{equation} \tilde {q}(t_0) = q_0 ,\ \ \  \tilde {p}(t_1) = p_1
\label{(2.3)} \end{equation}

\begin{equation}\gamma_2   : \tau  \rightarrow  \bigl( \tilde {\tilde {q}}(\tau
),\tilde {\tilde {p}}(\tau )\bigr) ,\ \  {\rm for} \ \  t_1\leq \tau  \leq t
\end{equation}

\begin{equation} \tilde {\tilde {p}}(t_1) = p_1 ,\ \ \   \tilde {\tilde {q}}(t)
= q         \label{(2.4)} \end{equation}

\noindent We define

               \begin{equation} (qt\vert p_1t_1)   = (2\pi \hbar )^{-n/2}
               \sqrt{D_{++}} \exp [ i S_{++}(qt,p_1t_1)/
               \hbar] \label{(2.5)} \end{equation}

\begin{equation} (p_1t_1\vert q_0t_0) = (2\pi\hbar )^{-n/2} \sqrt{ D_{--}}\exp
[
i S_{--}(p_1t_1,q_0t_0)/\hbar ]               \label{(2.6)} \end{equation}
where

\begin{equation} D_{++}= det \left(\frac{ \partial^2 S_{++}}{ \partial q^i
\partial p_{1j}}\right) \label{(2.7) }\end{equation}

\begin{equation} D_{--} = det \left( \frac{\partial ^2 S_{--}}{ \partial q^i_0
\partial p_{1j}} \right)      \end{equation}

\noindent and $ S_{++} , S_{--}  $ are Legendre transforms of the classical
action along the two trajectories,

\begin{equation} S_{++}(qt,p_1t_1)  =  p_{1i} \stackrel{\approx} {q}_1^{i} +
\int_{\gamma_2} ( \tilde {\tilde {p}}_{i} d \stackrel{\approx} {q}^i(\tau )-
H(\stackrel{\approx} {q}(\tau ),\stackrel{\approx}{p}(\tau))) d\tau
\label{(2.9)}
\end{equation}

\begin{equation} S_{--}(q_0t_0,p_1t_1) = -p_{1i} \tilde {q}_1^{i} +
\int_{\gamma_1}(\tilde {p}_i d\tilde {q}^i- H(\tilde {q}(\tau),\tilde
{p}(\tau)))\  d\tau  \label{(2.10)} \end{equation}

\noindent where

\begin{equation}
\tilde {q}_1 \equiv   \tilde {q}(t_1), \ \ \ \   \ \stackrel{\approx} { q}_1
\equiv \stackrel{\approx}   {q}(t_1)
\label{(2.11)}
\end{equation}

\noindent
With the above notation we now introduce the definition of the short time
propagator (STP) $ (qt\Vert q_0t_0) $ by

\begin{equation} (qt\Vert q_0t_0) = \frac{1} {\sqrt{\rho (q)\rho (q_0)}} \int
d^n p_1 ~ (qt\vert p_1t_1) ~ (p_1t_1\vert q_0 t_0) \label{(2.12)}
\end{equation}

In  (\ref{(2.5)}) to (\ref{(2.10)}) it is understood  that  the  independent
variables   are  those  explicitly  shown on the left hand side. It is  also
understood   that  that the right hand sides have been expressed in terms of
the  independent variables using the equations (\ref{(2.3)}) , (\ref{(2.4)})
for  the  classical  trajectories $ \gamma_1$ and $ \gamma_2 $. The function
$S_{++}$  is  the  generator  of  canonical  transformation connecting `old'
co-ordinates  and  momenta  at  time  t to the  `new'  ones  at time $ t_1.$
Similarly,  $  S_{--}  $   is  the  generator  of  canonical  transformation
connecting  the  conjugate  variables   for the trajectory  $ \gamma_1  $ at
time  $ t_1 $ to those at time $ t_0.$ It must be emphasisized that $ S_{\pm
\pm } $  are need for short times only.

\

We shall define two path integrals,  the  first  one  is  without  any scaling
of time and  the  second  one  with  scaling  of  time.  The {\it first
hamiltonian path integral}  $ K[H,\rho ] $ defined  as  a  summation  over
 histories with $ (q\epsilon  \Vert q_00) $ inserted for short times:

\begin{equation}
K[H,\rho ](qt;q_0t_0) \stackrel {\rm def}{\equiv} \lim_{N \rightarrow \infty}
\int\prod^{N-1}_{k=1} \rho (q_k)dq_k \int\prod^{N-1}_{k=0} (q_{k+1}\epsilon
 \Vert q_k 0)   \hspace{.5cm}   \label{(2.13)} \end{equation}
with  $ \epsilon = t/N $   and $ q_N \equiv q$.

\
Motivated by the importance of the local scaling of time in the path integral
formalism, we now introduce the second hamiltonian path integral which will be
described as hamiltonian path integral with scaling of time. The definition
makes use
of hamiltonian path integral HPI1. A special case of the hamiltonian path
integral
with scaling was found suitable for quantization in arbitrary
coordinates$^{17}.$

Given a hamiltonian function  $H(q,p)$,  measure $ \rho (q)$,  and  a  strictly
positive function $ \alpha (q) $, we introduce {\it second hamiltonian path
integral with local scaling of time}, to be denoted by  $  {\cal K} [H,\rho
,\alpha ] $  as

\begin{eqnarray}
\lefteqn{ {\cal K} [ H, \rho , \alpha  ]( q t; q_0 0 ) }  \nonumber      \\
          & \equiv &  \sqrt{
          \alpha (q) \alpha (q_0)} \int \frac{dE}{(2\pi \hbar)} \, \exp(-
          iEt/ \hbar ) \int^{\infty}_0  d\sigma  K[\alpha (H-E),\rho ]
          (q\sigma ;q_00) \nonumber \\
          && \hspace{4 in} \cdots
          \label{(2.14)}
 \end{eqnarray}

\noindent

The right hand side of (\ref{(2.14)}) has the first hamiltonian path integral
with $
\alpha (q)(H-E)$ as `hamiltonian' function.{\it  The two  hamiltonian path
integrals
$K$ and $ {\cal K}$ will henceforth be called HPI1 and HPI2 respectively }.
This
completes the definition of HPI2. We shall now give several remarks concerning
the
definintion of HPI2 which will be helpful in clarifying our approach.
\begin {enumerate}

     {\item Our first remark concerns the definition of HPI2 and use of local
scaling
     of time in this paper.Although the definition of HPI2 is inspired by the
     expression and formulas appearing in the literature, we do not utilize or
rely
     on any of the other details, mathematical or otherwise, from papers in the
     existing literature on local scaling of time. }

     {\item The second remark is about realtionship between HPI1 and HPI2. The
path
     integral HPI2 generalizes HPI1 and conversely HPI1 becomes a special case
of
     HPI2 in the sense that for trivial scaling, when the scaling function
$\alpha
     (q) $ is a constant independent of time, HPI2 becomes equal to HPI1. For
this
     reason we shall continue to describe HPI2 as the hamiltonian path integral
with
     scaling. }

     {\item  The  third remark is on the appearance of the functions $\alpha
     (q)$  and $\rho (q)$ in the definition of HPI2.As far as the definition
     is  concerned  the  HPI2  is  defined  for arbitrary scaling function $
     \alpha  (q) $ and measure $ \rho (q) $ just as the hamiltonian function
     $  H(q,p)  $  is kept arbitrary in the definition of any path integral.
     Specific values such as those given by $\alpha (q)=\sqrt{g} $ and $\rho
     (q)=\sqrt{g}  $  give  rise  to  the  path integral scheme suitable for
     quantization in arbitrary coordinates ( see Sec. 3.2 below). }

     {\item  Our  last  reamark  concerns the properties and applications of
     HPI2.  The  results  on  scaling  developed  earlier$^{18}$  and others
     obtained  here  have been applied to exact path integration by relating
     the  path  integral for the problem to be solved to another problem for
     which answers are already known or can be worked out. When one attempts
     to  do  this  one  has  to,  in  general,  deal with other intermediate
     expressions  involving  HPI1 or HPI2 with different choices of $ \alpha
     (q) $ and $ \rho(q). $ It is for this purpose $ \alpha (q) $ and $ \rho
     (q) $ have to be kept arbitrary while defining HPI2. }
\end{enumerate}
In  the  remaining  part  of  this  section  we  shall  summarize  important
properties of the two hamiltonian path integrals

\

{\it 2.2 Descretized form} :  In the limit $ \epsilon   \rightarrow 0$, the
only
important terms are are the terms of order $ \epsilon $  and only these need be
retained in the  definition of STP. Retaining such terms only, the canonical
STP
is equivalent to
           \begin{eqnarray}
 (q\epsilon  \Vert q_00)   = \left(\rho (q)\rho
   (q_0)\right)^{-1/2} \int\frac{d^n p}{(2 \pi \hbar )^n}
    \exp \left[ ip_i(q^i-q^i_0)/ \hbar \right] \times \hspace{1.0 in}  \cr
\exp \left[ -i \epsilon \bigl(H(q,p) + H(q_0,p)\bigr) / \hbar \right]
\left[ 1 -\frac{\epsilon }{2} \frac{\partial ^2H(q,p)}{\partial q^i \partial
 p_i} +\frac{\epsilon}{2}\frac {\partial ^2 H(q_0,p)}{\partial
                   q^i_0 \partial p_i} \right]  \label{(2.15)}
                                        \end{eqnarray}
\noindent
Inserting (\ref{(2.15)}) in (\ref{(2.13)})

  \begin{equation}
     K[H,\rho ](qt;q_0t_0)  =  \lim _{N \rightarrow \infty} \int
     \left( \prod^{N-1}_{k=1}\rho (q_k)dq_k \right)  \prod_{m=0}^{N-1}
     (q_{m+1}\epsilon  \Vert q_m 0)  \label{(2.16)}
  \end{equation}
\noindent
gives  discrete  form  for  the first hamiltonian path integral. Notice that
products  over  $ k $ and over $m$ appearing in ({\ref{(2.16)}) have different
ranges. Hence this  expression  involves $ N-1 $ fold $ q$ integrations  and
$   N-  $  fold $  p $ integrations. It is easy to do the $p -$ integrations
in  the  STP  if  the  hamiltonian  is  quadratic  in  momenta. This gives a
lagrangian form of STP which can be brought to a standard form making use of
the  McLaughlin  Schulman  trick. This makes a connection of our hamiltonian
path integrals with lagrangian form of path integral possible

\

{\it 2.3 Immediate consequences of definition of path integrals } :
We make some  useful  observations  which  follow  directly  from  the
definitions and the discrete form for the short time approximation.

 (a) Addition of a constant to hamiltonian $H$ is  equivalent to multiplying
 the HPI1 by a phase

\begin{equation}K[H-E,\rho ](qt;q_00) = \exp( iEt/\hbar ) K[H,\rho ](qt;q_00)
\label{(2.17)} . \end{equation}

(b) The dependence of the HPI1 and HPI2 on the measure $ \rho $ is particularly
simple. Thus for example we have

\begin{equation}\sqrt{\rho (q)\rho (q_0)} K[H,\rho ]  = K[H,1]  \label{(2.18)}
\end{equation}
\begin{equation}
K[H,\rho_1] = \sqrt{(\alpha (q)\alpha (q_0))} K[H,\rho _2]
\label{(2.19)} \end{equation}
\noindent
where $\rho_1  $ and $\rho_2  $ are related by \begin{equation}\rho _2(q) =
\alpha
(q) \rho _1(q). \end{equation}
\noindent
Similar relations are true for HPI2 $  {\cal K}$ also.

(c) When the scaling function $\alpha$ is a constant, say $ c $, we have

\begin{equation}{\cal K} [ H, \rho , \alpha =c] = K[ H, \rho ]
\label{(2.20)} \end{equation}
\noindent
This equation is equivalent to

 \begin{equation}\int \frac{dE}{(2\pi \hbar)} \exp(-iEt/\hbar )\int_0^{\infty}
d\sigma  K[c(H-E),\rho ](qt;q_0 0) = \frac{1}{c}  K[H,\rho ]( qt;q_00)
\label{(2.21)} \end{equation}

(d) In the special case when the function  $H(q,p) $ is independent  of  $q^j $
for some $j$ and depends on $p_j$ alone, the HPI1 depends  only  on  the
difference $(q-q_0)$, apart from an overall factor of $ \sqrt{\rho (q)\rho
(q_0)} $. This is easily seen  by doing the corresponding $p$ integrations
after
writing the discrete form  for HPI1.

\

{\it 2.4 Schrodinger equation for } $ {\cal K}$ :    If we take

\begin{equation}
H = \frac{1}{2m} g^{ij} p_ip_j+ V(q)  \label{(2.22)}
\end{equation}

\noindent
the HPI2 $ {\cal K}[H,\alpha,\rho](qt;q_0t_0)$ as  defined  in (\ref{(2.14)})
satisfies the Schrodinger equation
\begin{equation}
     i\hbar  \frac{\partial}{\partial t}  {\cal K}[H,\alpha,\rho]
     =(H_\alpha)_{op} {\cal K}[ H,\alpha,\rho] \label{(12.23)}
\end{equation}
with
     \begin{equation}
     (H_\alpha)_{op} = -\frac{\hbar^2}{2m}\rho^{-1/2}
          \alpha(q)^{-1/2}  \left(\frac {\partial}{\partial q^i}
          ( g^{ij}\alpha )\frac {\partial}{\partial q^j}
          \right) \rho^{+1/2} \alpha(q)^{-1/2}  + V(q)     \label{(12.24)}
\end{equation}
\noindent
and  $ {\cal K} $ has initial value given by
\begin{equation}
     \lim_{ t \rightarrow t_0}  {\cal K}[H,\sqrt{g},\sqrt{g}](qt;q_0t_0)
     = (1/\rho(q)) \delta (q-q_0)          \label{(12.25)}
\end{equation}

\noindent

For the special case when $ \alpha  = \rho  = \sqrt{g} $ the HPI2 $ {\cal
K}[H,\sqrt{g},\sqrt{g}](qt;q_0t_0)$  with  scaling,  as  defined  in
(\ref{(2.14)})
satisfies the Schrodinger equation

\begin{equation} i\hbar  \frac{\partial }{\partial t}  {\cal
K}[H,\sqrt{g},\sqrt{g}] =\widehat{H}{\cal K}[ H,\sqrt{g},\sqrt{g}]
\label{(2.23)}
\end{equation}
with

\begin{equation}\widehat{H}=
-\frac{\hbar^2}{2m}g^{-1/2}\frac{\partial}{\partial
q^i}( g^{ij} g^{1/2})\frac{\partial}{\partial q^j} + V(q)     \label{(2.24)}
\end{equation}
\noindent
and  $ {\cal K} $ has initial value given by
\begin{equation}
\lim_{ t \rightarrow t_0}  {\cal
K}[H,\sqrt{g},\sqrt{g}](qt;q_0t_0) = g^{-1/2}(q) \delta (q-q_0)
\label{(2.25)} \end{equation}

The relations (\ref{(12.23)}) to (\ref{(2.25)}) are the central results in our
scheme. In our  scheme $  {\cal K}[H,\sqrt{g},\sqrt{g}]$ is the  candidate for
the  quantum  mechanical propagator for a Hamiltonian of type (\ref{(2.22)})
because the right hand side  of (\ref{(2.24)}) can be recognized as the
Schrodinger operator $ - (\hbar^2/2m) \nabla^2+ V(q) $.

\

The proof of (\ref{(12.24)}) and (\ref{(12.25)}) is long and can be given in
several ways. Here we indicate, in brief, one of the methods used in ref. 18.
For
this purpose consider the "propagation of functions of q in time" governed by
the
HPI2 $ {\cal K}$ and define

     \begin{equation}
     \Psi  (q,t) = \int \! \rho (q_0) dq_0 \, {\cal K}
     [H,\rho,\alpha](qt;q_00) \Psi (q_0)          \label{(2.26)}
     \end{equation}
\noindent
The quantity $ \Psi (q,t+\Delta t) - \Psi (q,t)$ is computed up to the first
order  terms  for  small  $\Delta  t.$  To  do  this   we  by  first  insert
discrete   form   for  the  HPI1 $K[\alpha (H-E),\rho ]$  in the right  hand
side   of   (\ref{(2.14)})   and  use  the  resulting expression for $ {\cal
K}[H,\rho,\alpha(q)]  $  in the right hand side of (\ref{(2.26)}).   Because
in   (\ref{(2.14)})   integral   of  $ K[\alpha (H-E),\rho](q\sigma ,q_00) $
over  $  \sigma   $  is   needed,   it is not possible to replace $ K $ by a
single  STP and, we have to retain full discrete  expression for finite $N$.
The  expression  (\ref{(2.26)})  involves  $N $ fold  $q-  $ and  $N  $ fold
$p-$  integrations  as  well  as  integrations over $ E $ and $\sigma $. The
integration   over $ E $ results in a Dirac delta function implying a linear
constraint in $\sigma $ and  $\Delta t $ with functions of $q $ appearing as
coefficients.  The  $\sigma  -$  integration  is easily done using the delta
function.  The  resulting  expression  is  expanded  in powers of $\Delta t$
retaining  all   $O(\Delta  t) $ terms. The result of the remaining $N$ fold
$q-$ and $p- $integrations in (\ref{(2.26)}) can be expressed as an operator
acting  on $\Psi $ after long algebraic manipulations. he result can also be
proved  by  making use of the correspondance of the HPI1 with lagrangian form
of path integral mentioned earlier at the end of Sec. 2.2.

 As a by product of the above result we also get the scaling formula given
below.


{\it 2.5 Scaling formula} : The HPI2 $ {\cal K}[H,\rho ,\alpha ]$ with scaling
function $\alpha (q)$  and density $\rho $, as defined by (\ref{(2.14)}) above,
is
related to the HPI2 with  trivial scaling and hence to HPI1 with hamiltonian $H
- U(\alpha )  $
\begin{equation}{\cal K}[H,\rho ,\alpha ] =  {\cal K}[H-U(\alpha ),\rho ,1] =
K[H-U(\alpha ),\rho ]              \label{(2.27)} \end{equation}

\noindent
The precise form of $U(\alpha ) $ depends on $ H$ used. For $H(q,p) $ given by
(\ref{(2.22)}) we have
\begin{equation}U(\alpha ) =
-\frac{\hbar^2}{8m}\left({g^{ij}\partial_i(\ln\alpha
)\partial_j (\ln\alpha ) + 2 \partial_i(g^{ij}\partial_j (\ln \alpha ))}
\right)
 \label{(2.28)} \end{equation}
\noindent
Redefining  $  H  $  in  the  scaling  formula
(\ref{(2.27)}),  it  can be put into a useful form as a relation between two
HPI1.
           \begin{eqnarray}
           \lefteqn{  K[H,\rho ](qt;q_00)  } \nonumber  \\
           &=&  {\cal K} [H+U(\alpha ),\rho ,\alpha
          ](qt;q_00)  \nonumber \\
          &=& \sqrt{\alpha(q)\alpha(q_0)}\! \int\! \frac{dE}{(2 \pi \hbar) }
          \exp(-iEt/\hbar )\!
          \int_0^{\infty}\! d\sigma  K[\alpha (H-E+U(\alpha )),\rho
          ](q\sigma ;q_00) \nonumber \\
          && \hspace{4.5 in}   \cdots
                     \label{(2.29)} \end{eqnarray}

\noindent
This formula expresses HPI1  without  scaling  in  terms  of  a  HPI1  with
scaling.The factor $\sqrt{\alpha(q)\alpha(q_0)}$ can be absorbed into change of
measure  from $\rho  $ to $  \rho / \alpha  $ for the HPI1 in the right hand
side of
(\ref{(2.29)})


\begin{eqnarray}
K[H,\rho](qt;q_00)  \nonumber  \hspace{3.50in}
\end{eqnarray}
\begin{equation}
= \! \int \! \frac{dE}{(2 \pi \hbar) }\exp(-
iEt/\hbar )\! \int_0^{\infty}\!  d\sigma  K[\alpha (H-E+U(\alpha )),(\rho /
\alpha)](q\sigma ;q_00)  \label{(2.30)}
\end{equation}
\noindent
Other useful forms of the above relations are obtained by integrating  over
time
or taking inverse  Fourier  transforms.  As  an  example, we can obtain
\begin{eqnarray}
\int dt K[\alpha (H-E),\rho ](qt;q_00)  \nonumber \hspace{2.5in}
\end{eqnarray}
\begin{equation}
=  \frac{1}{\sqrt{\alpha(q)\alpha(q_0)}} \int_0^{\infty} dt K[(H-E+U(\alpha
)),\rho ](qt;q_00)  \label{(2.31)}
\end{equation}
by integrating the scaling relation  (\ref{(2.27)}) over time.

\

\begin{center}
{\bf 3. Point transformations and canonical path integration}
\end{center}
{\it 3.1 Initial  condition  for  canonical  path integration }:
For a path integral  to  give  rise  to  correct  propagator,  it  is
important that not only the correct Schrodinger equation be reproduced,  it
must
also satisfy  boundary  conditions  appropriate  to  the  co-ordinates chosen.
In this section we will at first take an illustrative  example  and then
discuss
the general case of initial condition for the  canonical  path integration. As
an example of initial condition  on  the  propagator  in  arbitrary
coordinates,
consider the change from the cartesian co-ordinates  to  polar co-ordinates in
a
plane.

\begin{equation}x= r \cos\theta ,\ \ \  y = r \sin \theta
\label{(3.1)}  \end{equation}

\noindent
The  right  hand  side  of  (\ref{(3.1)})  is  invariant,  separately,  under
the following transformations.

\begin{eqnarray} && (i)\ \ r \rightarrow \  r  , \ \ \ \  \theta \rightarrow
\theta
+
2m \pi ;              \label{(3.2)} \\
&& (ii) \  r \rightarrow  - r , \ \ \  \theta \rightarrow  \theta  + (2m+1)
\label{(3.3)}  \end{eqnarray}

\noindent
Thus we have the relation

\begin{equation}
\delta (x-x_0) \delta (y-y_0) = \\
\frac{1}{r} \sum_{m=-\infty}^{\infty} \bigl( \delta (r-r_0) \delta (\theta
-\theta _0
+ 2\pi m) + \delta (r+r_0) \delta [\theta -\theta _0 + (2m+1)\pi] \bigr)
\label{(3.4)}\end{equation}

Quantum mechanical  propagator, $ \langle \vec{r} \, t\vert \vec{r}_0 \,
t_0\rangle$,
in  two  dimensions satisfies the initial condition in cartesian co-ordinates

\begin{equation} \lim_{t \rightarrow t_0} \langle \vec{r}, t \vert
\vec{r}_0t_0\rangle = \delta  ( \vec{r}-\vec{r}_0)
\label{(3.5)}\end{equation}
\noindent
Therefore, in plane polar  co-ordinates  the  propagator,  at  equal  times,
reduces to the right hand side of  (\ref{(3.4)}).  Conversely  for  the  change
of variable from  $ r,\theta   $ to  $ x,y  $

\begin{equation}r = (x^2+ y^2)^{1/2}, \ \ \ \ \ \  \theta  = \tan^{-1}(y/x)
 ,
\label{(3.6)}\end{equation}

\noindent
let us assume that a hamiltonian path integral, say HPI1, is constructed to
satisfy the initial condition

\begin{equation}K [H,r](r\theta  t_0;r_0\theta _0t_0 ) =\frac{1}{r} \delta (r-
r_0)\delta (\theta -\theta _0)               \label{(3.7)} \end{equation}

\noindent
Because the  right hand side of  (\ref{(3.6)})  does  not  change  when   $
(x,y)
\rightarrow (-x,-y) $ , therefore, the equation

\begin{equation}\frac{1}{r} \delta (r-r_0) \delta (\theta -\theta _0) = \delta
(x-
x_0) \delta (y-y_0) +  \delta (x+x_0) \delta (y+y_0)     \label{(3.8)}
\end{equation}

\noindent
correctly gives the boundary condition (\ref{(3.7)}) in the cartesian co-
ordinates. In general let  $ q,p \rightarrow  Q,P  $  be a point
transformation.

\begin{equation}
Q=Q(q)  ,\ \ \ \ \ \ \  P=P(q,p)               \label{(3.9)}\end{equation}

\noindent
with the inverse transformations written as

\begin{equation}q=q(Q),\ \ \ \ \ \ \ p = p(Q,P)   \label{(3.10)}\end{equation}
\noindent
Let $ h(q,p) $  and  $ H(Q,P)=h(q(Q),p(Q,P)) $ be hamiltonian functions in the
two sets of co-ordinates. Let us set up HPI2 in the two sets of co-ordinates $
q
$  and  $ Q $  directly  using  the scaled canonical path integrals using the
same  classical  Hamiltonian  but expressed in terms of the appropriate
canonical  variables. The two  HPI2 ${\cal K}[h,\sqrt{g}_1,\sqrt{g}_1](qt;q_00)
$  and  $  {\cal K}[H,\sqrt{g}_2,\sqrt{g}_2](Qt;Q_00)  $  satisfy the same
Schrodinger equation and by construction they obey the following boundary
conditions in the limit  $ t \rightarrow  0 .  $

\begin{equation}
{\cal K}[h,\sqrt{g}_1,\sqrt{g}_1](q t=0;q_00) = g_1^{1/2}(q)\delta (q-q_0)
\label{(3.11)}
\end{equation}

\begin{equation}
{\cal K}[H,\sqrt{g}_2,\sqrt{g}_2](Q t=0;Q_00) = g^{1/2}_2(Q) \delta (Q-Q_0)
\label{(3.12)}
\end{equation}
\noindent
Whenever  the  function  $ q(Q) $ can be solved for $ Q $ uniquely, $  {\cal
K}[H,\sqrt{g}_2,\sqrt{g}_2]   $     will   be   obtained   from    $   {\cal
K}[h,\sqrt{g}_1,\sqrt{g}_1] $ of (\ref{(3.11)}) by expressing $ q $ in terms
of  $Q.$  This  relationship will not hold whenever the relation  $ q=q(Q) $
cannot  be  inverted  to give $ Q $ uniquely in terms of $q.$ This is due to
the  fact  that  in  such  a case the right hand sides of (\ref{(3.11)}) and
(\ref{(3.12)}) are not equal.
\noindent
Let  $ \Lambda $  be a transformation  on  $ Q's $  such that  $ q(\Lambda
Q)=q(Q) $ . The set of all such transformations form  a group. Following
Moshinsky et al. we  call  it ambiguity group of the transformation  $ q
\rightarrow Q,  $ and denote it by $ {\cal A}.$ Then

\begin{equation}
     \delta (q-q_0) = \vert J \vert \sum _{\Lambda \in  {\cal A}} \delta
     (Q-\Lambda Q_0)           \label{(3.13)}
          \end{equation}
\noindent
In the above equation  $ Q_0  $ stands for any one solution of the  $ q=q(Q)  $
and  $ J  $ is the Jacobian of the transformation  $ q\rightarrow  Q $ . The
relation between the two propagators  (\ref{(3.11)}) and (\ref{(3.12)}) can the
be written down
\begin{eqnarray}
\lefteqn  { {\cal K}[h,\sqrt{g_1},\sqrt{g_1}](qt;q_0 0) \hspace{4.5 in}}
\nonumber \\
&& =\sum _{\Lambda \in  {\cal A}}  {\cal
K}[H,\sqrt{g_2},\sqrt{g_2}](Qt;\Lambda Q_0 0)=\sum _{\Lambda \in  {\cal A}}
 {\cal K} [H,\sqrt{g_2},\sqrt{g_2}](\Lambda
Qt;Q_0 0) \       \label{(3.14)}
\end{eqnarray}
\noindent
In the equation (\ref{(3.14)}) it is understood that the variables  $ q, q_0  $
are to  be expressed  in  terms  of   $ Q  $ and  $ Q_0   $  by  using  the
equations  for the point transformation. The result (\ref{(3.14)}) follows from
the fact that both sides obey the same Schrodinger equation and the same
boundary condition.


{\it 3.2 Quantization in arbitrary co-ordinates:} We  are now in  a  position
to write the path integral for quantization in arbitrary  co-ordinates  needed
for later applications. We are interested  in  the  potential  problems  of
quantum mechanics. The classical hamiltonian, $ h_{cl},  $  for these problems
is of the form

\begin{equation}h_{cl} = \vec{p}^{~ 2}/ 2m + V(\vec{r})
\label{(3.15)}
\end{equation}
\noindent
where $  \vec{r}  $  denotes cartesian co-ordinates and  $ \vec{p}  $ are the
conjugate momenta. Let  $ Q  $ be any other set of co-ordinates related to  $
\vec{r}  $ by a point transformation . Then  the  quantum  mechanical
propagator  $  \langle  \vec{r} t\vert \vec{r}_0 0  \rangle    $ satisfying
the Schrodinger equation and the correct boundary condition
\begin{equation}\lim_{t \rightarrow 0} \ \langle \vec{r}t\vert \vec{r}_0 0
\rangle
=   \delta (\vec{r}-\vec{r}_0)             \label{(3.16)}\end{equation}
\noindent
can be written in terms of HPI2 set up in $ Q, P $ variables as
\begin{eqnarray}
\lefteqn{ \langle \vec{r} t\vert \vec{r}_0 0\rangle } \nonumber \\
 & = &\sum _{\Lambda \in
       {\cal A}} {\cal  K}[H_{cl},\sqrt{g},\sqrt{g}](\Lambda Qt;Q_00)
      \nonumber      \\
& = & (g(Q)g(Q_0))^{1/4}\sum _{\Lambda \in {\cal A}} \int\frac{dE}{(2 \pi
\hbar)}
            \exp(-iE t/\hbar ) \int_0^{\infty} d\sigma K[H_E,
            \sqrt{g}] (\Lambda Q \sigma ;Q_00) \nonumber \\
& = & \sum _{\Lambda \in  {\cal A}}  \int \frac{dE}{(2 \pi \hbar) }
        \exp(-iE t/\hbar ) \int_0^{\infty}  d\sigma  K[ H_E ,
        \rho =1] (\Lambda Q \sigma ;Q_0 0)  \label{(3.17)}
\end{eqnarray}

\noindent
where $ H_E =\sqrt{g}(H_{cl}-E) $. The corresponding energy dependent
Green function is defined by
\begin{equation}{\bf  G}(\vec{r}t,\vec{r}_0 \vert E) = \int_0^{\infty}  dt
\exp(iE_+
t/ \hbar) ~ \langle \vec{r}t \vert \vec{r}_0 0 \rangle ,
\label{(3.18)}\end{equation}

\noindent
where $E_{+}=E+ i \eta  , \eta > 0 $ and $ \lim {\eta \rightarrow 0 } $ is
taken at the end. For the energy dependent Green function
we have the following path integral representation in arbitrary co-ordinates.

\begin{eqnarray}
\lefteqn { {\bf G}(\vec{r},\vec{r}_0 \vert E) \hspace{4.0 in} } \nonumber \\
&=& (g(Q)g(Q_0))^{1/4}\sum _{\Lambda \in{\cal A}} \int_0^{\infty} d\sigma
K[\sqrt{g}(H_{cl}-E),\sqrt{g}](\Lambda Q \sigma Q_00)      \label{(3.19)} \\
&=& \sum _{\Lambda \in  {\cal A}}  \int_0^{\infty}  d\sigma  K[\sqrt{g}(H_{cl}-
E),\rho =1](\Lambda Q \sigma ,Q_0 0)             .  \label{(3.20)}
\end{eqnarray}
\noindent
In  the  above  $ H_{cl}  $ is the classical hamiltonian written in terms of
$  Q  $  and  $ P $ . {\it The expressions (\ref{(3.17)}) and (\ref{(3.20)})
define,  respectively,  the  scheme  for  obtaining  the  quantum mechanical
propagator  and  the  energy  dependent  Green  function.  They  involve the
classical  Hamiltonian only which is to be expressed in terms of appropriate
co-ordinates and  conjugate  momenta, no ad hoc addition of  $  O(\hbar^2) $
terms is necessary in any co-ordinate system. }

  \begin{center}

{\bf 4.  Relating path integrals }
           \end{center}
\vspace{5 mm}
In  many  of  the exact evaluations of the path integrals in literature, the
results  are  be  obtained  by  relating the desired proppagators to one for
which  answers are known. There are some notable investigations in which the
authors actually compute the multi-dimensional integrations to complete path
integration.  Besides the use of local rescaling of time, the  technique  of
adding  new degrees of freedom with known, preferably trivial,  dynamics has
proved  to  be  useful in relating path integral representations for quantum
mechanical propagators of different quantum  mechanical problems. Adding new
degrees  of  freedom  is  like  inverse  of  separating the variables in the
differential equation approach. We shall briefly recall this technique here.

Let  $ H_{ext}(q,p,P)  $ be independent of   $ Q  $  and   $ H(q,p)  $  be
such that  for  $ P= $  constant, say  $ \nu  $ , we have

          \begin{equation}H_{ext}(q,p,P)\vert_{p=\nu}       = H(q,p)
          \label{4.1)}\end{equation}
          \noindent
We  will say that $ H_{ext} $ is an extension of $ H $ by addition of a new
degree   freedom  $ Q,P  $ with trivial dynamics. Let  $ K $  and  $ K^{ext}
$ be  the two HPI1 using  the  unit measure  and with  $ H  $ and  $ H_{ext}
$ as hamiltonians. Then

          \begin{equation}
          K[H,1](qt;q_00) = \int d(\Delta Q)  \exp(i\nu \Delta Q)
          K[H_{ext},1](qQt;q_0Q_00)
                \label{(4.2)}\end{equation}

          \noindent
Notice that $ H_{ext} $ is assumed to be independent of $ Q.$ It, therefore,
follows that  $ K[H_{ext},1]  $  depends only on  $ Q-Q_0 (\equiv \Delta  Q)
$   .  The  limits  of  integration  over  $  \Delta Q  $ are usually from $
-\infty  $ to $ +\infty. $ However, if  $ K[H_{ext},1]  $  is periodic in  $
\Delta  Q  $  the  region of integration can be restricted to one period for
the  case  $  \nu  = $ integer. Similar results have been used in literature
and  one  can  easily  arrived at the result by considering the differential
equation  satisfied by the the two path integrals appearing in (\ref{(4.2)}).

In  this paper almost all the examples of path integrations to be discuused,
with  the  execption of the hydrogen atom in the parabolic co-ordinates, the
desired  propapgator or the energy dependent Green function will be obtained
by  relating the corresponding  path integral representation to another path
integral  which is known. A careful distinction has to be maintained between
the   path  integrals  which  are  related to quantum mechanical answers for
propagator of some problem and other path integrals which appear only in the
intermediate   steps.  For  any quantum mechanical problem the propagator is
given   by  (a)  the  corresponding  HPI1  in  cartesian  co-ordinates or an
equivalent   HPI2   expression   with   trivial   scaling, or (b) expression
(\ref{(3.17)})   appropriate    to  the  non-cartesian  co-ordinate  system
selected,  or  (c) the expression in the  line  just  before (\ref{(3.17)}).

Apart  from the addition of new degrees of freedom, properties about of HPI1
and  HPI2  when  $\rho $ or $ \alpha $ are changed will be used in obtaining
relations between different path integrals.

A  few  remrks about the notation to be followed are given in the following.
We  shall  always  use  the  bold  face symbols $ {\bf K}_0 \langle  qt\vert
q_0t_0\rangle    $   and    $   {\bf  G_0}  (q,q_0\vert E), $  with  a  zero
subscript,to   denote  the  full {\it quantum mechanical  solution}  for the
propagator   and  the   energy   dependent   Green  function,  respectively.
Similarly,  the hamiltonian  for  the  problem  to  be  solved  will have  a
subscript $ 0. $

Different superscripts or subscripts on roman, script or  bold face  symbols
will   be   used  to  denote  other  intermediate  or  related expressions .
Many   of  the  exact  treatments  in   this paper  begin  with  a series of
observations  and  from  these observations we "zero in" to  one  of the two
required  functions   $  {\bf K}_0\langle qt\vert q_0t_0\rangle $ or  $ {\bf
G}_0(q,q_0\vert E).  $

In the following sections the symbol ${\bf k_1} \langle xt\vert x_00\rangle$
will  be  used  to denote harmonic oscillator propagator in one dimension for
mass $ \mu $ and frequency $ \omega $ .
          \begin{equation}
          {\bf k}^{osc}_1 \langle xt\vert x_00\rangle =\frac{\mu
          \omega}{2  \pi  i
          \sin \omega t }  \exp\left( i \frac{\mu  \omega}{2 \sin \omega t
          } \bigl[(x^2+
          x^2_0) \cos \omega t - 2 x x_0 \bigr] \right)
          \end{equation}

          \noindent
The  propagators for harmonic oscillator in n dimension will be product of n
such  factors  and will be denoted by  $  {\bf k}^{osc}_n   $ . Also in  the
following we shall use same symbol to denote the classical hamiltonian  in
different co-ordinates.
\
\vspace{3mm}
\begin{center}
{\bf  5.  Potential problems in one dimension }
\end{center}

\

{\it 5.1  The  $ ax^2 + b/x^2  $  potential on half line}: The solution
for  this problem will be obtained by relating it, following  ${\rm
Duru}^{18}$,
to two  dimensional oscillator problem in a half plane. The
derivation is written as a sequence of  steps  consisting   of   several
small   intermediate  results   and observations.

\

           {\it Step1}  : The hamiltonian for the problem given by

\begin{equation}
H_0  =\frac{p^2}{2m} + ax^2 + b/x^2    .                  \label{(5.1)}
\end{equation}
\noindent
The required propagator  $ {\bf K}_0  \langle  x t \vert x_0 t_0\rangle  $
has initial value

\begin{equation}
\lim _{t \rightarrow t_0} {\bf K_0} \langle xt \vert x_0t_0 \rangle  =
\delta (x -
x_0)                 \label{(5.2)}
\end{equation}
\noindent
The propagator  $ {\bf K}_0 $ , in terms of the hamiltonian path integral
representation, is given by
\begin{equation}
{\bf K}_0\langle  x t\vert x_0 0\rangle =  {\cal K} [H_0,\rho =1, \alpha
=1](xt;x_0t) = K [H_0,\rho =1](xt;x_00)        \label{(5.3)}
\end{equation}
This is due to the fact that HPI1 in (\ref{(5.3)}), the  last  expression,
is  the right  answer  in  the  Cartesian  co-ordinates.  It  obeys   the
correct Schrodinger equation and the correct initial condition at  $ t=0. $

\

{\it  Step 2} : The classical hamiltonian  $ H_0  $ looks like the
hamiltonian $  H_1  $

\begin{equation}
H_1 =  \frac{p^2_x}{2m}+\frac{p^2_{\theta}}{2m x^2}  + ax^2
\label{(5.4)}
\end{equation}
\noindent
with  $ p_{\theta}  $ taken as suitable constant. Under a point
transformation

\begin{equation}
x_1 = x \cosh \theta    ,\ \ \    x_2 = x \sinh \theta
\label{(5.5)}
\end{equation}

\begin{equation}
dx_1 dx_2  = x dx d\theta  ,\ \ \   \sqrt{g}  =  x         \label{(5.6)}
\end{equation}
\noindent
the hamiltonian  $ H_1  $ takes the form of harmonic oscillator
hamiltonian in two dimensions

\begin{equation}
H_1   = \frac{p^2_1}{2m} -\frac{p^2_2}{2m}  +\frac{1}{2} m\omega ^2(x^2_1 -
x^2_2)               \label{(5.7)}
\end{equation}
\noindent
for which the the exact quantum mechanical propagator is known.

\

{\it Step 3  } : We set up the path  integral  quantization  scheme  for
oscillator problem in the two sets of co-ordinate  systems,  viz.,  the
plane  polar co-ordinates  $ (x,\theta )  $ and cartesian co-ordinates
$ \vec{x} = (x_1, x_2)$ and define

\begin{equation}
     {\cal   K}_1 \equiv  {\cal K} [ H_1 ,1,1]( \vec{x},t ; \vec{x}_0 0)
     =       K [H_1 ,1](\vec{x},t ; \vec{x} 0)          \label{(5.8)}
\end{equation}
\noindent
and
\begin{eqnarray}
{\cal K}_2
&\equiv& {\cal K} [ H_1 ,\sqrt{g}, \sqrt{g}](x,\theta t ;x_0\theta _00)
 \hspace{2.0in} \\
&=& \int\frac{dE}{(2 \pi \hbar) } \exp(- iE t/\hbar ) \int d\sigma
K [H_2, \sqrt{g}](x,\theta  \sigma  ;x_0\theta _0 0)  \label{(5.9)}
\end{eqnarray}
\noindent
where
\noindent
\begin{equation}
H_2 = x(H_1 -E) = x\bigl( \frac{p^2_x}{2m}+\frac{p^2_{\theta}}{2m x^2} + ax^2
- E \bigr)      \label{(5.10)}
\end{equation}
\noindent
The two functions  $  {\cal K}_1  $ and  $  {\cal K}_2  $ become equal when
expressed in terms of  either co-ordinates because the mapping  $x_1, x_2 $
 to   $ x,  \theta  $  is  one  to  one.  They are also equal to the quantum
mechanical propagator in two dimensions for (\ref{(5.7)}).

\begin{equation}
{\cal K}_1  =   {\cal K}_2  ={\bf k}_1^{osc}\langle  x_1t \vert
x_{10}0\rangle {\bf k}_1^{osc}\langle x_2t \vert x_{20} 0\rangle
\label{(5.11)}
\end{equation}
\noindent
 $  {\bf k}^{osc}_1\langle  x t\vert x_0 0\rangle  $ appearing in the above
equation is  the  propagator  for  one dimensional oscillator with mass  $
m  $ and frequency
 $  \omega    $ given  by   $ \sqrt{2a/m}.  $

\

{\it Step 4} : Although  $ H_1   $ reduces to $  H_0  $ for  $ p_{\theta} =
$ const. but  $ x  $  dependences  of the normalizations of  $  {\cal K}_2
$ and  $  {\bf K_0}  $ do not match:
\noindent
\begin{equation}
{\bf K}_0 (x,t=0;x_0,0) =  \delta (x-x_0)
\end{equation}
\noindent
and
\begin{equation}
            {\cal K}_2(x,\theta ,t=0;x_0,\theta _0,0) =\frac{1}{x} \delta
            (x-
            x_0)\delta (\theta -\theta _0)
\end{equation}
\noindent
The steps to relate these two have to take into account of this difference
in the initial conditions at  $  t=0.  $  In order to have  correct
initial  value for the two dimensional problem, we define
\begin{equation}
\tilde{\bf K}(xt \vert x_0t_0)   =  \frac{1}{\sqrt{xx_0}}   {\bf  K}_0
\langle xt\vert x_0t_0\rangle                 \label{(5.12)}
\end{equation}
which obeys
\begin{equation}
\tilde{\bf K} (xt\vert x_0t_0) \vert _{t=t_0}=\frac{1}{x}   \delta (x-x_0)
\label{(5.13)}
\end{equation}

\

{\it Step 5} : Next We insert the HPI1 of (\ref{(5.3)}) for  $ {\bf K}_0
\langle
xt;x_0 0 \rangle  $ in the right  hand side of (\ref{(5.12)}) and use the
formula
(\ref{(2.31)}). Dividing (\ref{(2.29)}) by $ \sqrt{\alpha(q)\alpha(q_0)} $ and
using it we get
\begin{eqnarray}
\lefteqn{\tilde{\bf K}\langle  xt\vert x_0 0 \rangle}  \nonumber \\
\ \ \ & = & \frac{1}{\sqrt{x x_0}}
K[H_0,\rho =1]  \nonumber \\
\ \ \ & = & \int dE \exp(-iEt / \hbar ) \int^{\infty}_0
d\sigma  K[H_3 ,\rho =1](x,\sigma ; x_0 0)  \label{(5.14)}
\end{eqnarray}
\noindent
where
\begin{eqnarray}
H_3 \nonumber  &=&  x(H_0+ U(x) - E ) \nonumber   \cr
&=&  x \left\{ \frac{p^2}{2m} + \left( b + \frac{\hbar^2}{2m} \right)
\frac{1}{x^2} + ax_2 - E \right \}             \label{(5.15)}
\end{eqnarray}

\

{\it Step 6} : This problem of obtaining HPI1 for $  H_3   $ in (\ref{(5.13)})
is
related  to the HPI1 for the extended hamiltonian  $ H_2  $

\begin{equation}
H_2 = x \left( \frac{p^2_x}{2m} +\frac{p^2_{\theta}}{2m x^2}  + ax^2   - E
\right)                      \label{(5.16)}
\end{equation}
\noindent
for  $ p^2_{\theta}   =  \left(2mb/ \hbar  + 1/4 \right)  $
.  Thus using (4.2)  we get the relation
\noindent
\begin{equation}
K[H_3 ,\rho =1](x,\sigma ; x_00) = \int d(\Delta \theta ) \exp(i\nu \Delta
\theta ) K[H_2,\rho =\sqrt{g}](x\theta \sigma ; x_0\theta _00)
\label{(5.17)}
\end{equation}
\noindent
where   $ \Delta \theta  = \theta -\theta _0  $ and
\begin{equation}
 \nu ^2  =  \left( \frac{2mb}{\hbar^2}   +\frac{1}{4}  \right)
\label{(5.18)}
\end{equation}

\

{\it Step 7} : From (\ref{(5.12)})  $  {\bf K}_0   $  is given in terms of
$  \tilde{\bf K}_0  $ which can be  expressed  in terms of  $ K[H_2,1]  $
using (\ref{(5.14)}) and (\ref{(5.17)}). Thus we obtain
\begin{eqnarray}
\lefteqn{{\bf K}_0  \langle  xt\vert x_0,0\rangle} \nonumber  \\
 \ &=& \sqrt{xx_0}   \tilde{{\bf K}} <xt \vert x_00>  \nonumber \\
 \ &= &\sqrt{xx_0}  \int dE \exp(-iEt/ \hbar ) \int^{\infty}_0 d\sigma
K[H_3 ,\rho =1](x,\sigma ; x_0 0) \nonumber  \\
 \ &=&  \int  d(\Delta  \sigma  )  \exp(i\nu   \theta   )   \int   dE
\exp(-iEt/ \hbar ) \int d\sigma   K[H_2, x ](x\theta \sigma
;x_0\theta _00)  \nonumber   \\
 \ &= & \int d(\Delta \sigma ) \exp(i\nu \theta )  {\bf k}^{osc}_2
\label{(5.19)}
\end{eqnarray}
\noindent
Here  the  last  step comes from equality (\ref{(5.11)}) already noted and $
{\bf  k}^{osc}_2  $  is  the  propagator  for  the  two  dimensional problem
\ref{(5.7)}).We  get  the  desired  result  by and following ref. 19 for the
remaining steps and integrations. Omitting the details, the final answer is
\begin{eqnarray}
     {\bf K}_0 \langle  x t\vert x_00 \rangle
     = \frac{m \omega \sqrt{xx_0}}{\hbar
     \sin\omega t}  I_{\nu} \left(\frac{m \omega \sqrt{ x x_0}}{i \sin
     \omega t} \right)
     \exp \left[ \left(\frac{i m \omega}{2\hbar} \right) \cot \omega t
     (x^2 + x^2_0 ) \right]          \label{(5.20)}
\end{eqnarray}

\

{\it 5.2 Morse Oscillator} : We shall briefly indicate how   the
propagator  for this problem is related to that for  $ ax^2 + b/x^2  $
potential on half line.  The Morse oscillator hamiltonian
\begin{equation}
H_0  = \frac{p^2}{2m} + A \exp(-2\alpha x) -B \exp(-\alpha x)
\label{(5.21)}
\end{equation}

\noindent
after a point transformation to a new coordinate  $ r = exp(-ax/2) $
becomes
\begin{equation}
H_0   = \frac{1}{2m}a^2 r^2p^2_r + ( A r^4 - B r^2)          \label{(5.22)}
\end{equation}

\noindent
Quantization of the Morse oscillator in the new variable  $ r  $ requires
us  to set up
\begin{eqnarray}
\lefteqn{{\cal K}[H_0 ,\sqrt{g},\sqrt{g}](rt;r_00)}   \cr
&\equiv& \sqrt{\alpha(q)\alpha(q_0)} \int \frac{dE}{(2 \pi \hbar) } exp(-
iEt/ \hbar ) \int_{0}^{\infty} d\sigma  K[ H_E, \rho ](r\sigma ;r_00)
\label{(5.23)}
\end{eqnarray}

\noindent
where  $  \alpha  =\sqrt{g}= (2/ar)  $ and  $ H_E  $  is given by
\begin{eqnarray}
 H_E & =&   (2/ar) ( H_0 -E)         \label{(5.24)}   \cr
\ &=& (ar/2) \left( \frac{p^2_r}{2m}  + \frac{4}{a^2}(Ar^2- B)
-\frac{4E}{a^2r^2}\right)       \label{(5.25)}             \cr
\ &\equiv & (ar/2) H_2
\end{eqnarray}

          \noindent
where $ H_2 $ is the function in the brackets in the second line of equation
(\ref{(5.25)}).  The HPI1 in r.h.s. of (\ref{(5.23)}) can be related to HPI1
for  $  H_2  $ using the scaling formula with a $ U(ar/2) $ appearing in the
Hamiltonian.  The  HPI1  for the  resulting hamiltonian $ H_2 - U(ar/2) $ is
just the HPI1 for the radial oscillator and is known from the results in the
previous subsection.


\begin{center}
{\bf                          6. Coulomb problem in two dimensions}
\end{center}
The Coulomb problem in two dimensions is solved easily by  our  method in the
parabolic co-ordinates as it gets related to harmonic oscillator  in two
dimensions. The steps leading to the solution are as follows.

           Step 1: The classical hamiltonian for the Coulomb problem in two
           dimensions is
           \begin{equation}
                  H_0 = \frac{\vec{p}^{ ~ 2}}{2m} -\frac{e^2}{r}
                  \label{(6.1)}
                                                              \end{equation}
                                                            \begin{equation}
              \vec{p} = (p_x,p_y)    ,\vec{r} = (x,y)
\label{(6.2)}
                                                              \end{equation}
\noindent
In the  " 2-dimensional parabolic coordinates "  $ u_1, u_2   $ defined
\begin{equation}
x  =  u_1^2  - u_2^2; \   \   \  \   y  =  2u_1 u_2           \label{(6.3)}
\end{equation}

\noindent
           the classical hamiltonian takes the form
           \begin{equation}
H_0 =  \frac{\vec{p}_u^{ ~ 2}}{8m\vec{u}^2} -\frac{e^2}{\vec{u}^2}
                                                  \label{(6.4)}
\end{equation}
\begin{equation}
dx dy  =  4\vec{u}^{ ~ 2} du_1 du_2, \ \ \  \  g^{1/2} = 4 \vec{u}^2
\label{(6.5)}
\end{equation}

{\it Step 2 :} For quantization of the Coulomb system in  the  cartesian,
recall for cartesian co-ordinates the desired propagator   $ {\bf K}_0 \langle
\vec{r} t\vert \vec{r}_0 0 \rangle   $ is  just  the HPI2 with a trivial
scaling, which is also  equal  to  HPI1  for  the  same hamiltonian,  Thus we
have

\begin{equation}
{\bf K}_0 \langle \vec{r} t \vert \vec{r} 0\rangle =   {\cal K}[H_0 ,1,1] =
K[H_0 ,1]
\label{(6.6)}
\end{equation}

{\it Step 3:} Quantization in parabolic coordinates  $ u_1,u_2   $ will proceed
 via
path integral HPI2 with hamiltonian $ H_0,$  scaling function  $ \alpha  =
4u^2.   $
Thus
we should set up $   {\cal K} [H_0 ,4u^2, 4u^2]( \vec{u}t;\vec{u}_0 0)  $
           \begin{equation}
{\cal K}[H_0 ,4u^2,4u^2](\vec{u}t; \vec{u}_0 0) = \int \frac{dE}{(2 \pi \hbar)}
\exp(-iEt/\hbar ) \int_{0}^{\infty} d\sigma  K[ H_E, 1](\vec{u} \sigma
;\vec{u}_0 0)
                                                             \label{(6.7)}
                                                               \end{equation}
\noindent
with

\begin{equation}
           H_E  =  4u^2 (H_0  - E)   = \left(\frac{\vec{p}^{ ~2}_u}{2m}- 4e^2 -
           4E \vec{u}^{ ~ 2} \right)
                               \label{(6.8)}
                               \end{equation}
\noindent
Notice  that  apart  from an additive constant, $ H_E $ is just the harmonic
oscillator  hamiltonian  in  two  dimensions.  Thus  we see that the HPI1 in
(\ref{(6.7)})   is  related  to  the  propagator  for  the  two  dimensional
oscillator  and the relation is given by
     \begin{equation}
           K[H_E,\rho =1]( \vec{u} \sigma ;\vec{u}_0 0) = \exp(i4e^2\sigma
           / \hbar ) {\bf k}^{osc}\langle u_1\sigma \vert u_{10}0 \rangle
           {\bf k}^{osc} \langle u_2 \sigma \vert u_{20} 0 \rangle
           \label{(6.9)}
                            \end{equation}
\noindent
with  $ {\bf k}^{osc}  $   is the one dimensional oscillator propagator for
mass   $ \mu =m  $ and frequency  $ \omega ^2= -8E/M.  $

{\it Step 4 :} The HPI2  ${\cal K}$  of (\ref{(6.7)}) is thus known. We use
analysis of Sec. 3  to relate HPI2 for $ H_0 $ in the cartesian and the
parabolic coordinates. For this  we notice that $  x,y  $ defined by
(\ref{(6.3)}) do not change if we substitute

\

\begin{equation}
u_1 \rightarrow - u_1,  \ \ \ {\rm and} \ \ \  u_2  \rightarrow   - u_2 .
\label{(6.10)}
\end{equation}
\noindent
Therefore, we have

\begin{equation}
            \delta (\vec{r} -\vec{r}_0)  =  g^{-1/2} \left[ \delta  (\vec{u}-
            \vec{u}_0) +
                            \delta  (\vec{u}+\vec{u}_0) \right]
\label{(6.11)}
\end{equation}

\noindent
and
\begin{equation}
      {\cal K}[H_0 ,1,1](\vec{r}t;\vec{r}_0 0) = \bigl[ {\cal K} [H_0 ,4u^2,
      4u^2](\vec{u} t; \vec{u}_0 0) + {\cal K} [H_0 ,4u^2, 4u^2](\vec{u} t;
      -\vec{u}_0 0) \bigr]  \label{(6.12)}
\end{equation}

{\it Step 5:} Thus the desired quantum mechanical propagator $ {\bf K}_0 $ is
obtained from (\ref{(6.6)}) and (\ref{(6.12)}) using the expressions
(\ref{(6.7)})
and (\ref{(6.9)}). Writing the  answer in terms of the energy dependent Green
function we get

\begin{eqnarray}
{\bf G}_0 (\vec{r},\vec{r}_0\vert E)
= \int_{0}^{\infty} \exp(4ie^2\sigma /\hbar )\bigl[ {\bf k}_1^{osc}\langle u_1
 \sigma \vert u_{10}0 \rangle {\bf k}_1^{osc} \langle u_2
\sigma \vert u_{20}0 \rangle    \cr
 +   {\bf k}_1^{osc}\langle u_1 \sigma \vert -u_{ 10}0\rangle  {\bf
 k}_1^{osc}\langle u_2 \sigma \vert -u_2 0\rangle \bigr]
 \label{(6.13)}
\end{eqnarray}

\noindent
This  result agrees with equation (28)-(30) of ref. 20 after the integration
variable  $ \sigma   $ is replaced by  $ \tau  = 4\sigma .  $


\begin{center}
                            {\bf 7.  H-Atom in three dimensions}
                                                        \end{center}
\

\noindent
{\it      7.1. K-S Transformation: }  The H-atom in  three  dimensions  is
related  to  isotropic harmonic oscillator   in   4-dimensions   via   K-S
transformation.   Local  time transformations in the path integral formalism
and K-S transformation  have been used to relate  the  Coulomb  problem
Green  function  to  the  Green function for the oscillator  problem  in
4-dimensions.  This  relation  is derived in our canonical formalism of path
integrations follows.

\

{\it Step 1:} Consider the 4-dimensional oscillator problem defined by
           \begin{equation}
              H_4^{osc}  = \frac{\vec{p}^{~2}_u}{2M} + \frac{1}{2} M \Omega ^2
\vec{u}^{~ 2}                  \label{(7.1)}
                      \end{equation}
           where
           \begin{eqnarray}
                \vec{u} & =&  (u_1, u_2 , u_3, u_4)
\label{(7.2)} \\
                           \vec{p}_u & =&  (p_{u_1},p_{u_2},p_{u_3},p_{u_4})
                \label{(7.3)}
                                                              \end{eqnarray}
           We introduce the new co-ordinates  $ (x_1,x_2,x_3,x_4) $ , in two
steps. In the first step the transformation equations
           \begin{eqnarray}
  u_1 & =& \sqrt{r}\sin\frac{\theta}{2} \cos (\frac{\alpha+\phi}{2}) \ \ \ \
  ;\ \ \ \     u_2    =     \sqrt{r}\cos     \frac{\theta     }{2}     \sin
(\frac{\alpha-\phi}{2})
\cr
  u_3    & =& \sqrt{r}\cos\frac{\theta}{2}   \cos  (\frac{\alpha-\phi}{2}) \ \
\ \
    ; \ \ \ \         u_4  \   =   \sqrt{r}\sin   \frac{\theta}{2}    \sin
(\frac{\alpha+\phi}{2})
\label{(7.4)}
\end{eqnarray}
define  the  co-ordinates  $  r,\theta ,\phi ,\alpha $ . In the next step we
identify  $  x_4=\alpha $ and the equations relating $ r,\theta ,\phi $ to $
(x_1,x_2,x_3)  \equiv \vec{x} $ are taken to be the same as those giving the
relation  between  the  spherical  polar  to cartesian co-ordinates in three
dimensions. Thus

           \begin{equation}
           x_1 = r \sin\theta  \cos\phi,\ \ \  x_2 =r \sin\theta \sin\phi,
           \ \  x_3 = r \cos\theta, \ \  x_4= \alpha       \label{(7.5)}
                                     \end{equation}
The  classical  hamiltonian (\ref{(7.1)}) when expressed in terms of the co-
ordinates   $  (\vec{x},  \alpha  )   $  and  conjugate momenta  $ (\vec{p},
p_\alpha )  $ becomes

\begin{equation}
               H^{osc}_4  =  4r\left\{ \frac{\vec{p}^{~2}}{2M}  +
\frac{(x_1p_2-x_2p_1) z p_\alpha } {M r (x_1^2 + x_2^2 )}\right\} + \frac{1}{2}
M\Omega ^2r
                 \label{(7.6)}
             \end{equation}
           Also

           \begin{eqnarray}
           d^4 u  &=& 16 r \sin \theta \, dr \, d\theta \, d\phi \, d\alpha
\label{(7.7)}     \cr
           d^4 u  &=& (16/r) dx_1 dx_2 dx_3 d\alpha
\label{(7.8)}
           \end{eqnarray}

           \begin{equation}
                    g^{1/2} = (16/r) \equiv J
          \end{equation}

\

 {\it Step 2:} Next we set up path integrals for the  4-dimensional
oscillator problem in two sets of co-ordinates systems. In the cartesian
co-ordinates  $\vec{u} $  the relevant HPI2 is equal to the HPI1 for the same
hamiltonian

           \begin{equation}
                      {\cal K} [H^{osc}_4, 1,1 ] = K[H^{osc}_4,1 ]
        \label{(7.10)}
                                     \end{equation}
           The  path integrals in (\ref{(7.10)}) at  $ t=0 $  become equal to
  $ \delta ^4(\vec{u}-\vec{u}_0).$  Thus  two HPI's in (\ref{(7.10)})
coincide with  the  exact  quantum  mechanical  propagator, $ {\bf k}^{osc}_4$
for the four dimensional oscillator problem with mass  $ M $  and
frequency  $ \Omega .  $ To quantize the oscillator (\ref{(7.1)}) in
coordinates
$ (\vec{x} ,\alpha )  $ we set  up  the HPI2  $  {\cal K} [H^{osc}_4, J,J](
\vec{x},\alpha ,t;\vec{x}_0,\alpha _0,0)  $ which is given by

 \begin{eqnarray}
\lefteqn{{\cal K} [H^{osc}_4, J, J]( \vec{x},\alpha ,t;\vec{x}_0,\alpha
_0,0)} \nonumber \\
             &=&16 \int \frac{d{\cal E}}{(2 \pi \hbar) } \exp(-iEt/\hbar  )
\int_{0}^{\infty} d\sigma  K[ H_{\cal E}, \rho =1]( \vec{x},\alpha ,\sigma
;\vec{x},\alpha _0 0)          \label{(7.11)}
             \end{eqnarray}
           Taking inverse Fourier transform this relation is written as

           \begin{eqnarray}
             \lefteqn{\int_{0}^{\infty} d\sigma  K[ H_{\cal E}, \rho
             =1](\vec{x},\alpha
,\sigma ;\vec{x},\alpha _0 0)}     \cr
           &=&\frac{1}{16} \int dt \exp(i{\cal E}t/\hbar )  {\cal K}
[H^{osc}_4,
J, J]( \vec{x},\alpha ,t;\vec{x}_0,\alpha _0,0)  \label{(7.12)}
           \end{eqnarray}
The hamiltonian function appearing in the HPI1 of r.h.s of (\ref{(7.11)}) is
given by

           \begin{eqnarray}
                       H_{\cal E} &=&  J (H^{osc}_4 - {\cal E}) /16
                  \label{(7.13)}   \\
             &=& 4 \left\{ \frac{\vec{p}^{~2}}{2M}  + \frac{(x_1p_2-x_2p_1) z
p_\alpha }{M r (x_1^2 + x_2^2 )}
                 + \frac{1}{2} M\Omega ^2- {\cal E}/r \right\}
\label{(7.14)}
                                                              \end{eqnarray}
           A factor of 16 appears in (\ref{(7.11)}) to (\ref{(7.13)}) because a
trivial  scaling  of time has been included in (\ref{(7.11)})

\

{\it Step 3:} Next we note that two HPI2 set  up  for  the  oscillator
problem, appearing in (\ref{(7.10)}) and (\ref{(7.11)}), are normalized
differently. Their values at  $ t=0  $ are related by

           \begin{equation}
 (1/J) \delta ^4(\vec{x}-\vec{x}_0)\delta (\alpha -\alpha _0)
= [ \delta ^4(\vec{u}-\vec{u}_0) + \delta ^4(\vec{u}+\vec{u}_0) ]
         \label{(7.15)}
                                     \end{equation}
           Thus we must have

           \begin{eqnarray}
\lefteqn{{\cal K} [H^{osc}_4, J, J] ( \vec{x},\alpha ,t;\vec{x}_0,\alpha_0,0)}
                             \nonumber         \\
&=&  {\cal K} [H^{osc}_4, 1,1 ]( \vec{u}t; \vec{u}_0 0) +
           {\cal K} [H^{osc}_4, 1,1 ]( \vec{u}t;- \vec{u}_0 0)   \nonumber  \\
&=&  {\bf k}^{osc}_4\langle  \vec{u} t\vert  \vec{u}_0 0\rangle
          + {\bf k}^{osc}_4\langle  \vec{u} t\vert -\vec{u}_00\rangle
                    \label{(7.16)}
\end{eqnarray}
\noindent
where the oscillator propagator,  $ {\bf k}^{osc}_4  $ appearing in
the above  equation  is the four dimensional oscillator propagator.

\

           {\it Step 4:} We next set up path integral quantization of  Coulomb
hamiltonian in cartesian co-ordinates. The required energy dependent Green
function  is given by (\ref{(3.20)}) in arbitrary co-ordinates.  In  cartesian
co-ordinates  we have

           \begin{equation}
             {\bf G}_0(\vec{x},\vec{x}_0\vert E) = \int_{0}^{\infty} d\sigma
K[H_0-E, \rho =1](\vec{x} \sigma ; \vec{x}_0 0)              \label{(7.17)}
                                     \end{equation}
           where  $ H_0  $ is the Coulomb hamiltonian

           \begin{equation}
                              H_0 =\frac{\vec{p}^{~2}}{2m}  -\frac{e^2}{r}
                                                               \label{(7.18)}
                                       \end{equation}
\

           {\it Step 5:} We can now assemble the above results to obtain the
H-atom  Green function. It should be noticed that for  $ p_\alpha =0  $ the
function  $ H_{\cal E}  $ is the Coulomb hamiltonian apart from an additive
constant  $ \frac{1}{2} M\Omega ^2  $ .Therefore,  the  integral of  $ K
[H^{\cal E},1]( \vec{x} ,\alpha \, t; \vec{x}_0,\alpha _0 \, 0)  $ over $
\Delta \alpha  (\equiv \alpha -\alpha _0) $  is related to  the   HPI1  for
the function  $ H_0- E  $

           \begin{equation}
              K[H_0-E, \rho =1](\vec{x} t; \vec{x}_00) =  \int^{4\pi}_0
d(\Delta \alpha ) K [H_{\cal E},1](\vec{x} ,\alpha  t; \vec{x}_0,\alpha _0
0)      \label{(7.19)}
              \end{equation}
where we have identified  $ m=M/4, {\cal E} = -M\Omega ^2/2 $ and $ ^2={\cal
E}  $  . The integral  of  the left hand side of (\ref{(7.19)}) over  $ t  $
is  just  the energy dependent Green function (\ref{(7.17)}) for the Coulomb
problem . Therefore using (\ref{(7.12)}) and (\ref{(7.14)})  we get,

           \begin{equation}
            {\bf G}_0(\vec{x},\vec{x}_0\vert  E) =\frac{1}{16} \int d\sigma
\exp(ie^2\sigma ) \int_{0}^{4 \pi} d(\Delta \alpha ){\cal K
}[H^{osc}_4,J,J](\vec{x}
,\alpha  \sigma; \vec{x}_0,\alpha _0 0)
\end{equation}
which   on  using (\ref{(7.16)}) gives

\begin{eqnarray}
 \lefteqn{ {\bf G}_0(\vec{x},\vec{x}_0\vert E) } \hspace{5.0 in} \cr
            = \frac{1}{16}\int^\infty _ 0
d\sigma \exp(ie^2\sigma )  \int_{0}^{4\pi} d(\Delta \alpha ) \bigl[
k^{osc}_4(\vec{u}\sigma ,-\vec{u}_0 0) + {\bf k}^{osc}_4(\vec{u}\sigma
,-\vec{u}_0 0)\bigr] \,     \label{(7.20)}
           \end{eqnarray}
           This  is  the  desired  connection  between  the  solutions  for
the  four dimensional oscillator and the Coulomb problems. The above result
(\ref{(7.20)}) is in agreement with the equation (108) of ref 19.

\

{\it  7.2 Parabolic Coordinates}: In this subsection we shall show
how exact path integration can be done for H-atom  in  parabolic  coordinates
within  the canonical formalism. In this case we proceed directly to set  up
the  path integral representation for the propagator and evaluate it.
                The coulomb hamiltonian

                \begin{equation}
H_{coul}  =\frac{\vec{p}^{~2}}{2m}  -\frac{e^2}{r}
                     \label{(7.21)}
                                                               \end{equation}
           takes the form

           \begin{equation}
H_{coul}  = \frac{p^2_\xi}{2m(\xi+\eta)}+\frac{p^2_\eta}{2m(\xi+\eta)}
                     +\frac{p^2_\phi}{2m\xi \eta}  -\frac{2e^2}{\xi^2+\eta^2}
      \label{(7.22)}
                                                               \end{equation}
           in the parabolic coordinates  $ \xi ,\eta ,\phi   $ defined by
           \begin{eqnarray}
                            x  &=&  \xi  \eta  \cos \phi  \cr
                            y   &=&   \xi   \eta    \sin   \phi
             \label{(7.23)}          \cr
                                   z  &=& \frac{1}{2}(\xi ^2 - \eta ^2)
                                   \end{eqnarray}
     Also

           \begin{equation}
dx\, dy\, dz \ \ = \ \  \xi  \eta  (\xi ^2+\eta ^2) d\xi\, d\eta \, d\phi
              \label{(7.24)}
                                                               \end{equation}
           We, therefore, set-up the HPI1  $ {\cal K} [H_E,\sqrt{g}]  $ in the
parabolic co-ordinates with

           \begin{eqnarray}
                     H_E  &=&  g^{1/2}(H_{coul} - E) \cr
   \                     &=& \xi  \eta  \left\{ \frac{p^2_\xi}{2m} +\frac{p^2_
   \eta}{2m}  +\frac{1}{2m} \left(\frac{1}{\xi^2}  +\frac{1}{\eta^2} \right)
    p^2_{\phi}  - E ( \xi^2 + \eta^2) - 2e^2  \right\}
\label{(7.25)}
\end{eqnarray}
           and use it to define HPI2 $ {\cal K}$ as follows

\[
 {\cal K} [H,\sqrt{g}\sqrt{g}] (\xi \eta\phi,t;\xi _0\eta_0
\phi_0,0)      \hspace{3.0 in}                  \nonumber \]
\begin{equation}
\ \ \   =  \int \frac{dE} {(2 \pi \hbar)} \exp(-iEt/\hbar  )
\int_{0}^{\infty}  d \sigma  K[H_E,1](\xi  \eta  \phi ,t ;\xi _0\eta
_0\phi 0,0)     \label{(7.26)}    \end{equation}
\noindent
$  {\cal K} $ satisfies the Schrodinger equation appropriate to the
Coulomb  potential but has the initial value
           \begin{equation}
           \lim_{t \rightarrow 0}    {\cal K}[H,\sqrt{g},\sqrt{g}](\xi ~ \eta ~
\phi
           ~ t; ~\xi _0 ~ \eta _0 ~ \phi _0 ~ 0) =
            \frac{1}{\xi \eta \left(\xi^2 + \eta^2 \right)}\delta (\xi -\xi
_0)\delta(\eta -\eta _0)\delta (\phi -\phi _0) \label{(7.27)}
                                                               \end{equation}
           whereas the Coulomb propagator should be normalized to

           \begin{equation}
                \delta (\vec{r}-\vec{r}_0)  =\frac{1}{\xi \eta \left(\xi^2 +
\eta^2 \right)} \delta (\xi  - \xi _0) \delta (\eta  - \eta
_0)\sum^{\infty}_{n=-\infty} \delta (\phi  - \phi _0 + 2\pi n)
\label{(7.28)}
           \end{equation}
           Taking (\ref{(7.28)}) it  into account the Coulomb propagator will
be
given by

\begin{eqnarray}
\lefteqn{{\bf  K}_0  \langle   \vec{r}   t\vert   \vec{r}_00\rangle }
                                                           \nonumber      \\
&=&\sum^{\infty}_{n=-\infty} {\cal K}[H_0,\sqrt{g},\sqrt{g}](\xi \eta \phi
+ 2 \pi n t ; \xi _0\eta _0\phi _0, 0)                                   \\
&=& \int \frac{dE}{(2 \pi \hbar) }\exp(-iEt/\hbar  ) \int_{0}^{\infty}
 d \sigma  \sum^{\infty}_{n=-\infty}  K[H_E,1](\xi \eta ,\phi
 + 2 \pi n,\sigma  ;\xi _0\eta _0\phi _0, 0)   \nonumber   \\
 \label{(7.29)}
           \end{eqnarray}
  Using time integrated version of (\ref{(2.31)}) and expressing change in
scaling as change in potential for HP1 we get, with   $ \alpha = \xi \eta  $ ,

           \begin{equation}
                     \int_{0}^{\infty} d\sigma  K[H_E,1](\xi \eta \phi  \sigma
 ; \xi _0\eta _0\phi _0 0)
                          =  \int_{0}^{\infty} K [h_E,1] (\xi \eta \phi
\sigma  ;\xi _0\eta _0\phi _0 0) d \sigma                    \label{(7.30)}
                          \end{equation}
      where

           \begin{equation}
                     h_E  = \frac{p^2_{\xi}}{2m} +\frac{p^2_\eta}{2m}
+\frac{p^2_{\phi}}{2m} \left(\frac{1}{\xi^2} + \frac{1}{\eta^2}\right) - E(\xi
^2 + \eta ^2) - 2e^2 -\frac{\hbar^2}{8m} \left(\frac{1}{\xi^2}  +
\frac{1}{\eta^2} \right)                                      \label{(7.31)}
                          \end{equation}
                          Thus we have

                          \begin{equation}
                   {\bf K}_0\langle \vec{r}t;\vec{r}_00\rangle  = \int
\frac{dE}{(2 \pi \hbar) } \exp(-iEt/\hbar ) {\bf G}_0(\vec{r},\vec{r}_0\vert E)
               \label{(7.32)}
                   \end{equation}
     with

                \begin{equation}
 {\bf G}_0(\vec{r},\vec{r}_0\vert E) =  \int_{0}^{\infty} d\sigma
\sum^{\infty}_{n=-\infty} K[h_E,1](\xi ,\eta ,\phi  + 2 \pi n,\sigma ;,
\xi_0, \eta _0, \phi _0, 0)    \label{(7.33)}
                                     \end{equation}
We shall now write the canonical path integral $  K[h_E,1]  $ in
the discrete  from and do the  $ \phi   $ and  $ p_\phi  $ path integration
explicitly. For this purpose we  write  $ K [h_E,1] $  of (\ref{(7.30)}) as

           \begin{equation} \int \prod_{j=1}^{N-1}  (d\xi _j d\eta _j)
           \left( \prod_{j=0}^{N-1} dp^\xi_j dp^\eta_j \right) exp
           \left[ \sum_{k} ip^\xi_k(\xi _{k+1}-\xi _k) + \sum_{k}
           ip^\eta_k( \eta _{k+1}-\eta _k)\right] \nonumber \end{equation}

                                    \begin{equation}
            \exp\left[ i \sum_{k} \frac{p_k^{\xi^2 }}{2m} +
\frac{p_k^{\eta^2}}{2m}- E(\xi ^2_k + \eta ^2_k) -\frac{\hbar^2}{8m}  \left[
\frac{1}{\xi^2_k} +\frac{1}{\eta^2_k}  \right]\right]   K_\phi
\label{(7.34)}
           \end{equation}
           where

           \begin{equation}
               K_\phi  =  \int \left(\prod^{N-1}_k d\phi _k \right)
 \left( \prod_{k=0}^{N-1}  dp^\phi_k \right)
                     \exp \left[ i  \sum_{j=0}^{N-1} \left\{ p^\phi_j (\phi
_{j+1} - \phi _j) + \frac{p_j^{\phi^2}}{2m} \left(\frac{1}{\xi^2_j}
+\frac{1}{\eta^2_j}  \right) \right\} \right]    \label{(7.35)}
                   \end{equation}
The  $ \phi _k  $ integral may first be done giving  rise  to  the
product  of  delta functions  $ \delta (p^\phi_k-p^\phi_{k+1})  $ which may in
turn be used to  do  all  except  one   $ p^\phi_k  $ integration. The last
remaining  $ p^\phi   $ integral  is  Gaussian  and  is  easily written down.
This gives

           \begin{equation}
                 K_\phi  =  \left( I\hbar  /2\pi i T \right)^{1/2} \exp \left(
                \frac{2iI(\phi-\phi_0)^2}{2T \hbar} \right)
\label{(7.36)}     \end{equation}
where

\begin{equation}
            I = \frac{1}{2m}\sum^{\infty}_{n=-\infty} \left( 1/\xi ^2_k +
1/\eta ^2_k \right) \Delta t               \label{(7.37)}
                                                 \end{equation}
           Using the identity

\[
\sum^{\infty}_{n=-\infty} \left( \frac{I \hbar}{2\pi i T}
 \right) ^{1/2}\exp \left( \frac{iI(\phi-\phi_0 +2\pi n)^2 }{2T\hbar} \right)
\hspace{2.5 in}  \nonumber \]
\begin{equation}
    =\sum^{\infty}_{n=-\infty} \left( \frac{1}{2\pi} \right)
\exp \left( -in^2T/2\hbar I \right) \exp \bigl[ in(\phi -\phi _0) \bigr]
       \label{(7.38)}
                    \end{equation}
           We get the Coulomb Green function as

           \begin{equation}
              {\bf G}_0(\vec{r},\vec{r}_0 \vert ~ E)  = \int_{0}^{\infty}
\sum^{\infty}_{n=-\infty} K [H_n,1](\xi,\eta,\sigma ;\xi _0,\eta_0,0)
\exp[in(\phi -\phi _0)] d\sigma           \label{(7.40)}
              \end{equation}
           where $  H_n $  in (\ref{(7.40)}) is the hamiltonian function

           \begin{eqnarray}
             H_n & = & \frac{p^2_{\xi}}{2m} +\frac{p^2_{\eta}}{2m} - E(\xi ^2 +
\eta ^2) +\frac{\hbar^2}{2m} (n^2 - \frac{1}{4})~(\frac{1}{\xi^2} +
\frac{1}{\eta^2}) \\
                   &\equiv &  \tilde{H}(\xi ,p_\xi) + \tilde{H}(\eta ,p_\eta)
                    \label{(7.41)}
                                     \end{eqnarray}
           As  $ H_n  $ is a sum of two  hamiltonians  each  depending  only
on  one  set  of conjugate variables, we have

           \begin{equation}
             K[H_n,1](\xi ~ \eta ~ \sigma ;\xi _0 ~ \eta _0 ~ 0)  =
K[\tilde{H},1](\xi  \sigma _0;\xi 0) K[\tilde{H},1](\eta  \sigma _0;\eta 0)
\label{(7.42)}
           \end{equation}
        The HPI1 for $ \tilde{H} $ is known from Sec. 5.1 and hence we get

           \begin{equation}
 K[H_n,1] =  \left( \frac{m \omega \sqrt{\xi \xi_0}}{i \hbar \sin
\omega t } \right)
       \left( \frac{m \omega \sqrt{\eta \eta_0}}{i\hbar \sin \omega t}\right)
        J_n \left( \frac{m \omega \sqrt{\xi \xi_0}}{i \hbar \sin \omega t }
                            \right)
        J_n \left( \frac{m \omega \sqrt{\eta \eta_0}}{i\hbar \sin \omega t}
         \right)   \nonumber                             \end{equation}

                           \begin{equation}
\exp \left\{\frac{im\omega}{2\hbar}  \cot \omega t \bigl(\xi ^2 + \xi
_0^2 + \eta ^2_0 + \eta ^2_0\bigr)\right\}         \label{(7.43)}
            \end{equation}

   \begin{equation}
            K  = \frac{1}{2\pi}\sum^{\infty}_{n=-\infty}  \exp[in(\phi -\phi
_0)] K[H_n,1](\xi,  \eta,  t; \xi _0, \eta _0,0)           \label{(7.44)}
                                     \end{equation}
The required Coulomb Green function is then given by

 \begin{eqnarray}
    \lefteqn{ {\bf G}_0 (\xi, \eta,  \phi ;\xi _0, \eta _0, \phi _0 \vert
E) } \nonumber    \\
  && = \int_{0}^{\infty} d\sigma  \exp(-2ie^2\sigma /\hbar)
\sum^{\infty}_{n=-\infty}  \left(\frac{m\omega}{i\hbar \sin \omega
\sigma}\right)^2
\exp(in(\phi -\phi_0))  \nonumber \\
  &&  \exp \left\{\frac{im\omega}{2\hbar}  \cot \omega\sigma
[ \xi^2 + \xi^2_0 + \eta^2 + \eta^2_0]\right\}
 J_n \left( \frac{m \omega \sqrt{ \xi \xi_0}}{\hbar \sin \omega \sigma}
\right)  \nonumber \\
&& J_n \left(\frac{m \omega \sqrt{ \eta \eta_0}}{\hbar \sin \omega
\sigma}  \right)                               \label{(7.45)}
\end{eqnarray}

\noindent
           Using the identity

           \begin{equation}
           \sum^{\infty}_{n=-\infty} \exp(in\alpha ) J_n(r)J_n(\rho ) = J_0(R)
                            \label{(7.46)}
                                     \end{equation}
        where

             \begin{equation}
                       R^2 = r^2 + \rho ^2 + 2r  \rho  cos \alpha
         \label{(7.47)}
           \end{equation}
           the Green function (\ref{(7.45)}) becomes

           \begin{eqnarray}
             \lefteqn{ {\bf G}_0(\xi, \eta,  \phi ;\xi _0,\eta _0,\phi _0\vert
E) }
             \\
                     &=& \int_{0}^{\infty} d\sigma  \exp(2ie^2\sigma /\hbar )
\left( \frac{m\omega}{i\hbar \sin \omega \sigma}\right)^2  J_0
\left(\frac{m\omega}{\hbar \sin \omega \sigma}  R\right)  \nonumber \\
&& \exp \left\{\frac{im\omega}{2\hbar} \cot \omega t (\xi^2 +\xi^2_0  + \eta^2
+\eta^2_0)\right\}            \label{(7.48)}
           \end{eqnarray}
           with R given by

           \begin{eqnarray}
                         R^2 &=& \xi \xi_0 + \eta \eta_0 + 2 \xi \xi_0
{}~ \eta \eta_0 ~ \cos (\phi -\phi _0)     \cr
                            & = &  r r_0 \cos (\psi /2)
     \label{(7.49)}          \end{eqnarray}
    and

           \begin{equation}
                       \cos \psi  = \cos \theta  \cos\theta _0 + \sin\theta
\sin\theta _0 \cos(\phi -\phi _0).               \label{(7.50)}
                       \end{equation}
           Defining  $ k = (m\omega /i\hbar ) $  the  above  result  is  seen
 to  coincide  with  that obtained by Ho and Inomata  in ref 13.

\

{\it  7.3 Radial propagator for Hydrogen atom }:  The hamiltonian for the
radial hydrogen atom problem is

           \begin{equation}
            H  = \frac{p^2 }{2m}- \frac{e^2}{r} +\frac{\hbar^2l(l+1) }{2mr^2}
          \label{(7.53)}
                                                               \end{equation}
Note that  $  1/r^2 $  term in the hamiltonian can be thought of as
coming from a  $  p_{\theta}^2/2mr^2   $ term   in   hamiltonian   for
corresponding   two   dimensional problem. Therefore, this  path integral
solution  of  this  problem  can  be obtained by relating it to the Coulomb
problem  in  two  dimensions.  This relation can be  derived  in a way
parallel to the derivation in  Sec.5.


\begin{center}
{\bf 8. Concluding Remarks}
\end{center}

We  have  summarized   the   main   properties   of   the  hamiltonian  path
integral formalism of ref 16, 17, and 18. We have also discussed application
to exact path integration. Here, we wish to highlight the following points.

(a)  For all other existing  literature  on  path  integration,  local scaling
of time is a technique used in exact solution, where as for us  it is an
essential ingredient of quantization scheme itself.  In  some  cases, such as
in
the example of H- atom in Sec. 6 , no further local  scaling  of time is
necessary to complete exact solution.

(b)  The  best  feature  of  this  approach  is  that  path   integral
quantization is  formulated  in  terms  of  the  classical  hamiltonian  in
arbitrary co-ordinates. Performing  changes  of  variables  becomes  quite
trivial in the path integral formalism in our scheme, as one can do this at the
classical level itself and use the classical hamiltonian  in  arbitrary
coordinates to  set  up  the  canonical  path  integral  quantization.  The
boundary conditions are correctly reproduced when initial condition for the
propagator is correctly taken into account as discussed in Sec 3. In  this
respect our approach is  new  and  more  direct  than  all  other  existing
treatments.

(c)  The canonical approach  offers  more  general  possibilities  for local
scaling of time, than those exploited here. It should be possible  to
generalize
the scaling relation to the case when $ f(q) $ is replaced by a function of
 $p $ or by a function of $p $ and $q$. It is not known  whether  this can be
used to do exact path integration of new problems.

(d)   A class of potentials, such as Scarf potential, Rosen  Morse potential
etc., can be related to a free particle on sphere. Though we are not able to
do  exact  path integration for free particle on a sphere, our scheme can be
used  to obtain inter-relations of propagators for the potentials within the
class.  A  possible approach to solve particle on sphere problem exactly may
be  to formulate it as a path integral for a constrained systems. For this a
discretized form of path integral for the constrained systems is needed.

(e)  In ref. 18 correspondence between the canonical and lagrangian forms of
path  integrals was established. For  most  of  the  problems,  not taken up
for   discussion   in   this   paper  one   can,   in   principle,   utilize
correspondence  between the hamiltonian and the  lagrangian  forms  of  path
integration  and  follow  the  existing  literature  to  complete  the  path
integration. However, this is not very interesting.

(f) Doing a close analysis of discussion in sections 5,6 and 7, we see that
with
exception of the path integral in  Sec.7.2,  hardly  any  path integral,  is
actually  computed.  All  the  details  of   computing   and manipulating path
integrals are buried in the results summarized in Sec.  2 which were derived in
 ref.
18. The  present   paper  thus  isolates  and identifies a set of properties of
Hamiltonian  path  integrals  useful  for path integration.

(g) The aim of path integration formalism is not merely to solve those quantum
mechanical problems  exactly  for  which  the  exact  solution  can already be
obtained by several methods. This  activity is useful only if it generates new
techniques or gives  new  insights  into  the  path  integral formalism. It
would be far more  interesting  and  useful  to  explore  the generalizations
and  advantages  of  the  hamiltonian  approach  for  path integration.

(h)  For  the constrained hamiltonian systems, in a most direct but a formal
approach,  one  sets  up path integral over the reduced phase in terms of an
independent  (reduced)  set  of  conjugate  variables  $(q*,  p*)$ which, in
general,  are  functions  of  an  initial  set $(q,p)$. By means of a formal
canonical  transformation,  when  the  path  integral  is  expressed as path
integral  over  the  initial set $(q,p) $ over full phase space, Dirac delta
functions  enforcing  constraints and determinants appear in  the integrand.
This  path  integral  requires a precise lattice definition. It then appears
that, any attempt to precisely define a discrete version of path integration
for  the  constrained  systems  must  solve  the  question  of incorporating
canonical  transformation  as  change  of  variables in the phase space path
integral approach. This merits further investigation.

{\it Acknowledgment:} One of the  authors (A.K.K.)  thanks  Prof.  H.S.
Mani  for hospitality extended to him at the Mehta Research Institute. We thank
Ramandip Singh Bhalla for help in preparation of the Latex version of the
manuscript.

\pagebreak

{\bf REFERENCES :}

\

\begin{enumerate}

\item Reviews and applications of path integration method can  be  found  in
several references. For example M.S. Marinov, Phys. Reports {\bf 60}, 1(1980)
{}.
D.C.  Khandekar  and  S.V. Lawande, Phys. Reports {\bf 137}, 115 (1986) .
A.S.Arthurs (ed.), Functional Integration and its Applications,  Proceedings
of
the  International Conference  at  London  1974,  Oxford  University  press
1975.   G.J. Papadopoulos  and  J.T.  Devrese  (eds.),  Path  Integrals  and
Their Applications to Quantum, Statistical and Solid State Physics,  Plenum,
New
York, 1978.  L.S. Schulman, Techniques and  Applications  of  Path Integration,
John Wiley, New York 1981. M.C. Gutzwiller,  A.  Inomata, J.R. Klauder, and L.
Streit (eds.), Path Integrals from  mev  To  Mev, World Scientific Singapore
1981.S.  Lundqvist  et  al.,  (eds.),  Path Integral Method and Applications,
Proceedings of Adriatico  Conference on  Path  Integrals,  Trieste,  Italy,
1-4
September   1987,   World Scientific, Singapore, 1988.

\item I.H. Duru and H. Kleinert, Phys. Lett {\bf 84B}, 185 (1979);  Fortschr.
der Phys. {\bf 30}, 401 (1982).

\item P. Kustaanheimo and E. Stiefel, J. Reine Angew Math.{\bf 218}, 204
(1965).
\item R. Ho and A. Inomata, Phys. Rev. Lett. {\bf 48}, 231 (1982); see also
      A. Inomata, Phys. Lett. {\bf 101A}, 253 (1984) ;
      F. Steiner, Phys. Lett. {\bf 106A}, 363 (1984).

\item P.Y. Cai, A. Inomata, and R. Wilson, Phys. Lett.{\bf 96A}, 117 (1983) .

\item A. Inomata and M. Kayed, J. Phys. {\bf A18}, L235 (1985).

\item A. Inomata and M. Kayed, Phys. Lett. {\bf 108A}, 9 (1983) .

\item M.V. Carpio and A. Inomata, in M.C. Gutzwiller, Path Integrals from
      mev  To  Mev,edited by A. Inomata, J.R. Klauder, and L. Streit (eds.),
      World Scientific Singapore 1981.

\item J.M. Cai, P.Y. Cai, and A. Inomata, Phys. Rev. {\bf A34}, 4621 (1986).

\item S.V. Lawande and K.V. Bhagat, Phys. Lett. {\bf 131A}, 8 (1988) ;
      D. Bauch, Nuovo Dim. {\bf 85B}, 118 (1985) .

\item W. Janke and H. Kleinert, Lett. Nouvo Cim. {\bf 25}, 297 (1979);  I.
Sokmen, Phys. Lett {\bf 106A}, 212 (1984) .

\item  A. Inomata, Phys. Lett. 87A, 387 (1982) ;H.Kleinert, Phys. Lett.
      {\bf 116A}, 201 (1986); I. Sokmen, Phys. Lett. {\bf 132A}, 65
     (1988); H. Durr and A. Inomata, J. Math. Phys. {\bf 26}, 2231 (1985).

\item N.K.Pak and I. Sokmen, Phys. Lett {\bf 103A}, 298 (1984);  Phys.  Rev.
{\bf A30}, 1629 (1984); I.H. Duru, Phys. Lett. {\bf 112A}, 421 (1985); I.
Sokmen, Phys. Lett. {\bf 115A},249 (1986); S. Erkoc and R. Sever,  Phys. Rev.
{\bf D30},  2117 (1984); I. Sokmen, Phys. Lett. {\bf 115A}, 6 (1986).

\item M. Moshinsky and T. Seligman in Lecture Notes in Mathematics vol
      {\bf 676} Springer Verlag (1978) and references therein.

\item B.S. DeWitt, Rev. Mod. Phys. {\bf 29}, 377 (1957).

\item A.K. Kapoor, Phys. Rev. {\bf D29}, 2339 (1984).

\item A.K. Kapoor, Phys. Rev. {\bf D30}, 1750 (1984).

\item A.K.Kapoor and Pankaj Sharan Hyderabad University  preprint
     HUTP-86/5 (1986).

\item I. Duru, ref. 13

\item I.H.Duru and H. Kleinert, Fortschr der  Physik {\bf 30}, 401 (1982)

\item  N.K Pak and I. Sokmen , ref. 13

\item L.D. Faddeev, Theor. Math. Phys. {\bf 30} ,3 (1969)

\item P. Senjanovic, Ann. Phys. (NY) {\bf 100}, 227 (1976)

\end{enumerate}
\end{document}